# Microbial Methanogenesis on Mars: A Life Detection Target for an MLE-Type Mission and Future Human Exploration

Rachel L. Harris (rachel.l.harris@nasa.gov)

***Disclaimer:*** *The views expressed in this white paper are those of the author and should not be construed to represent those of the Astrobiology Program, Science Mission Directorate, or NASA Headquarters.*

**Motivation.** Methane ($CH_4$) is one of the most enigmatic and potentially consequential trace gases detected on Mars, having demonstrated apparent seasonality[1], localized plumes[2–4], or, depending on who you ask, never conclusively detected at all[5,6]. While the Martian methane mystery remains unsolved, it has sparked a vibrant area of astrobiology research exploring whether microbial methanogenesis -- one of the simplest and most ancient metabolic pathways -- could plausibly proceed under various extreme physicochemical conditions characteristic of the Martian shallow subsurface. The emerging consensus is a cautious but compelling "possibly" (Table 1). These findings elevate methanogenesis as both a top priority candidate biosignature for extant life detection and a justification for urgently advancing in situ instrumentation capable of definitively detecting $CH_4$ and constraining its origin, biotic or otherwise.

**Table 1.** Traceability matrix of culturable methanogen tolerance to Mars-relevant environmental stressors and corresponding Mars Life Explorer (MLE)-type mission search environments. Methanogen "tolerance" is defined cumulatively: □□□ = no known tolerance; ■□□ = methane production; ■■□ = additional evidence of metabolic activity (e.g., transcriptional activity); ■■■ = growth and replication. Note: this table is not exhaustive and focuses on experimental studies from the Martian habitability literature to illustrate the range of experimental evidence supporting methanogenesis as a viable target for extant life detection. It does not include references to the broader methanogen or environmental microbiome literature, which certainly contain additional evidence of extremotolerance.

| Mars-Relevant Stressor | Tolerance Observed in Culturable Methanogens | Ref. | Corresponding MLE-Type Mission Search Environment |
|---|---|---|---|
| Desiccation | ■□□ | 7–11 | Salt crusts; regolith-air interfaces |
| Starvation | ■□□ | 12 | Oligotrophic regolith with low redox potential |
| Freeze-thaw cycling | ■■□ | 11–13 | Near-surface ice with diurnal/seasonal temperature cycling |
| Low-pressure | ■■■ | 9,13–15 | Everywhere |
| High salinity | ■■■ | 7 | Evaporites; brines |
| Acidic pH | ■■□ | 15 | Sulfate-rich deposits |
| Alkaline pH | ■□□ | 15,16 | Carbonate-bearing deposits; serpentinization-altered systems |
| Transient deliquescence of salts | ■■□ | 17 | Evaporites rich in hygroscopic/low-eutectic salts (e.g. $Ca(ClO_4)_2$, $Mg(ClO_4)_2$, $NaClO_4$, $CaCl_2$, $FeCl_3$, $FeCl_2$, $MgCl_2$,) |
| Perchlorates | ■■■ | 10,18–20 | Perchlorate-bearing regolith |
| Oxidative stress from free radical exposure | ■■□ | 21 | Reactive oxygen species (ROS) (e.g., $O_2^-$, $H_2O_2$, •OH) generated by UV photolysis and adsorbed onto regolith grains and dust |
| Ultraviolet (UV-A/B/C) radiation | ■ ■ □ (UV-C only) | 7 | Near-surface environments (< 5 mm depth)[22] |
| Solar Energetic Particles (episodic exposure) | □ □ □ | *no data available* | Near-surface environments (< 5 cm depth)[23] |
| Galactic Cosmic Rays (episodic exposure) | □ □ □ | *no data available* | Shallow subsurface (<3-5 m depth)[23,24] |

**Critical Technology Gaps.** If $CH_4$ is to be considered a top-priority candidate biosignature for life detection on Mars, we must equip missions with the tools necessary to measure it definitively. At present, no landed mission has demonstrated the ability to reliably detect $CH_4$ at sub-ppb levels, nor to distinguish its origin. This represents a critical technology gap. Future missions must be capable of detecting $CH_4$ and its isotopologues -- particularly $^{13}CH_4$ and $CH_3D$--with sufficient precision to determine $\delta^{13}C$ and $\delta D$ values that could support or rule out a biogenic source. This level of sensitivity will likely require a combination of tunable laser spectroscopy (e.g., CRDS or TLS) and pre-concentration via cold trap to enrich trace gases prior to measurement. As demonstrated by Chen et al.[25], dual-mode instruments that can alternate between isotopologue detection modes are particularly promising for *in situ* applications, though they remain in early stages of miniaturization and deployment readiness for Mars. Still, if this is to be a life detection mission, then enabling a definitive $CH_4$ detection capability must be treated as a mission-critical priority.

**Considerations of Low-Biomass and Low Metabolic Activity.** A persistent challenge in low-biomass, low-activity ecosystems is distinguishing true biological signal from background noise. On Mars, if extant life exists, its activity may occur at concentrations well below the detection thresholds of most instruments or biomarker kits. This raises the risk of false negatives, particularly in environments where energy limitation, desiccation, or cryogenic conditions suppress metabolism. One potential solution is supplementing passive detection with functional assays -- e.g., radiolabeled substrates like $^{14}C$- or $^{2}H$-bicarbonate -- to directly test for methanogenesis. Though this raises planetary protection concerns (and conjures the ghost of *Viking*), such techniques are familiar to researchers studying Mars analogs like the Atacama or deep biosphere and may provide unambiguous metabolic evidence in near-real time. These capabilities are tightly aligned with the goals of SFL-SAG, which aims to detect biosignatures and characterize habitability in the near-subsurface. A payload that includes methane isotopologue analysis and/or *in situ* metabolic assays would significantly advance the science return of an MLE-type mission, providing both context for biosignature interpretation and a means to distinguish between true biological activity and ambiguous abiotic signals. This is particularly vital in the post-*Viking* era, where confidence in detection claims demands a multi-pronged, converging-lines-of-evidence approach.

**Relevance to Human Exploration.** In addition to their scientific value for life detection, microbial methanogens may hold practical relevance for future human exploration. As summarized in Table 1, methanogenesis has been demonstrated under a range of simulated Martian stressors. Methane is a strong candidate for in situ resource utilization (ISRU) as a fuel, especially if microbial $CH_4$ production can be sustained using accessible $H_2$-rich subsurface brines and atmospheric $CO_2$ as feedstocks. Moreover, emerging research suggests some methanogens may play a role in perchlorate reduction[18,20], a critical challenge for ensuring astronaut safety given the toxicity of Martian regolith. While these applications remain speculative and require validation at scale, they highlight the potential versatility of methanogens not only as biosignature targets but also as potential biotechnological assets supporting long-term human presence on Mars.